# Kinetic modelling of the CO₂ capture and utilisation on NiRu-Ca/Al dual function material via parameter estimation

Meshkat Dolat[a], Andrew David Wright[b], Soudabeh Bahrami Gharamaleki[a], Loukia-Pantzechroula Merkouri[a], Melis S. Duyar[a,c], Michael Short[a,c*]

[a] School of Chemistry and Chemical Engineering, University of Surrey, Guildford, Surrey GU2 7XH, UK

[b] Department of Chemical Engineering, School of Engineering, The University of Manchester, UK

[c] Institute for Sustainability, University of Surrey, Guildford, Surrey GU2 7XH, UK

* m.short@surrey.ac.uk

**Abstract**

This study presents a detailed, open-source kinetic modelling computational framework for CO₂ capture and utilisation using a newly formulated dual-function material (DFM) comprising 15 wt% Ni, 1 wt% Ru, and 10 wt% CaO supported on spherical alumina. A finite difference reactor model was developed to simulate the cyclic adsorption, purge, and hydrogenation stages. The model incorporates experimentally-derived rate expressions, accounts for system delay via a second-order response function, and was fitted to time-resolved concentration laboratory data using Bayesian optimisation. A combined parameter estimation strategy was employed to ensure mass continuity across stages and improve the robustness of purge kinetics. The kinetic parameters extracted reveal that carbonate decomposition, not methanation, is the rate-limiting step during hydrogenation. Temperature-dependent simulations confirm a trade-off between reaction kinetics and CO₂ storage capacity, with methane yield maximised at 300 °C when compared with the other temperature sets. By offering transparent methodology and reproducible code, this work provides a robust platform for researchers and practitioners to study, validate, and optimise DFM systems.

## 1.    Introduction

Power-to-Gas (PtG) technology represents an important pathway for integrating renewable electricity into sectors where direct electrification is difficult, such as heavy industry, heating, and long-haul transport [1]. By converting excess electricity into synthetic natural gas (SNG) through hydrogen production and subsequent CO₂ methanation, PtG enables not only energy storage across seasonal timescales but also deep decarbonisation of existing gas infrastructures. This approach allows renewable energy to extend beyond the power grid, effectively embedding clean energy into the chemical and thermal economies [2]. However, conventional methanation processes require very pure CO₂ feeds, necessitating separate capture and purification steps that add cost and complexity [3]. Integrated carbon capture and utilisation (ICCU) aims to overcome these inefficiencies by combining capture and conversion into a single unit [4].

Dual-function materials (DFMs) have emerged as a promising class of sorbent-catalysts for ICCU. A DFM couples a basic CO₂ sorbent with a hydrogenation catalyst in one solid phase [5]. In practice, common sorbents include alkaline metal carbonates (e.g. Na₂CO₃, K₂CO₃) or oxides such as CaO and

MgO, while catalytically active metals are typically transition metals (Ni, Fe, Co) or precious metals (Ru) [6]. This combination enables reactive separation: the sorbent binds $CO_2$ from dilute streams, and the catalyst converts the bound $CO_2$ to methane when $H_2$ is introduced. DFMs can increase the energy efficiency of CCU by combining both steps into a single unit operation [7]. For example, CaO (a high-capacity carbonate-forming oxide) paired with Ru (a highly active methanation catalyst) is a widely studied system for capturing flue-gas $CO_2$ and converting it in situ to $CH_4$ [8].

DFMs operate in a cyclic mode consisting of three main steps: adsorption, purge, and hydrogenation (or reduction), as illustrated in Figure 1. During the adsorption step, a $CO_2$-containing feed (e.g., flue gas or air) flows over the DFM, and $CO_2$ chemisorbs onto the basic sorbent sites. This is followed by a purge step, in which an inert gas (typically $N_2$ or Ar) is introduced to remove residual gases and prevent gas-phase mixing in the subsequent hydrogenation stage. During hydrogenation, the chemisorbed $CO_2$ desorbs in the presence of $H_2$ and reacts to form $CH_4$ and $H_2O$. A second purge step is applied after hydrogenation to remove remaining products and prepare the system for the next adsorption cycle. These purge steps are essential not only to prevent unwanted side reactions, such as combustion from $O_2/H_2$ mixing, but also to enable accurate carbon balances by ensuring that only surface-bound $CO_2$ contributes to methane formation. By switching the gas phase between $CO_2$-rich and $H_2$-rich streams, DFMs eliminate the need for a separate thermal desorption step; hydrogenation serves both to release $CO_2$ from the sorbent and to convert it to fuel. Since all steps take place within a single reactor, $CO_2$ does not need to be transferred between separate units, and the high-temperature swing regeneration, typically required in amine scrubbing, is eliminated [9].

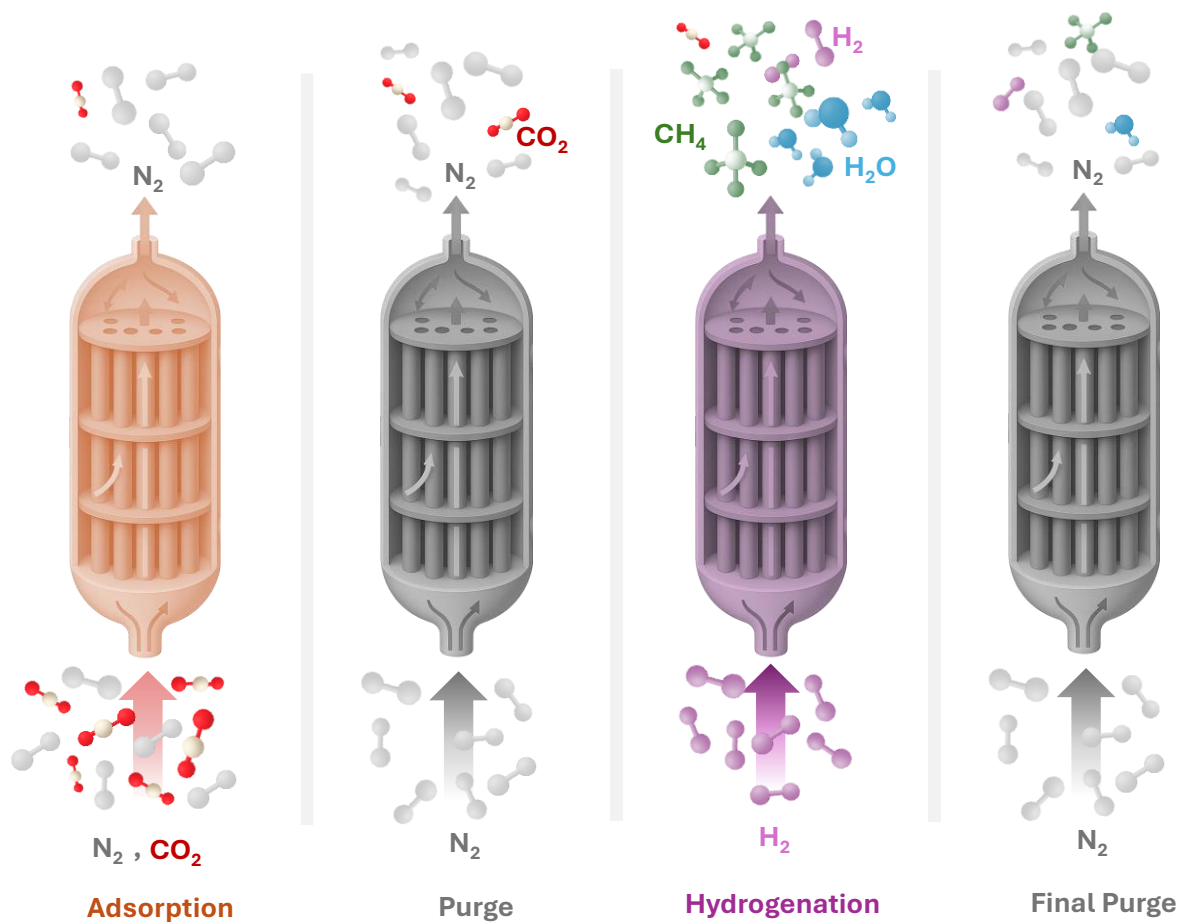

*Figure 1: Schematic of the cyclic operation of a DFM reactor, illustrating sequential adsorption, purge, hydrogenation, and post-reaction purge steps.*

The integrated DFM approach offers several process advantages. In situ capture and conversion greatly reduces the energy penalty of $CO_2$ capture, since no external heating is needed to regenerate the sorbent [10], [11]. Furthermore, because chemisorption on basic oxides has a high heat of adsorption, DFM systems can achieve high $CO_2$ affinity even from dilute streams [12].

Material design plays an important role in the performance of dual-function materials, and considerable research has focused on optimising formulation parameters such as sorbent-to-catalyst ratio, active phase dispersion, and the use of alkali or rare-earth dopants to enhance $CO_2$ capacity and catalytic reactivity [13], [14]. On the catalytic side, efforts have been made to improve the stability and activity of hydrogenation sites, with Ni and Ru widely studied for their respective trade-offs in cost, performance, and thermal stability [15], [16]. Various supports, including $Al_2O_3$, $SiO_2$, and $TiO_2$, have also been explored to promote metal dispersion and durability under cyclic operation [15].

While these material-focused studies have advanced the field considerably, they often operate independently of detailed process-level analysis. In particular, despite the growing body of experimental work on DFM formulations, there remains a notable lack of reliable kinetic models that capture their full cyclic behaviour and support scale-up and process optimisation. Such models are essential for informed process design and scale-up, as they determine key parameters like the required duration of the hydrogenation step, the rate of $CO_2$ uptake, and the expected methane yield per cycle. Critically, it is not sufficient for a DFM to exhibit high $CO_2$ capacity alone; if the associated reaction kinetics are too slow, significant $CO_2$ losses may occur during the purge or hydrogenation steps, or hydrogenation may need to be extended to impractical durations [10]. Therefore, predictive models that integrate both adsorption and purge dynamics and methanation kinetics are crucial for optimising reactor design and achieving efficient, compact, and high-throughput operation in ICCU systems.

To date, there are only few detailed report of kinetic modelling for a cyclic DFM process. One of the leading efforts was presented by Bermejo-López et al.[17], [18], who developed a dynamic reactor model for a 4%Ru–10%$Na_2CO_3$/$Al_2O_3$ DFM, formulating elementary adsorption and surface reactions to fit experimental gas profiles. Their coverage-based model could qualitatively reproduce $CO_2$, $CH_4$ and $H_2O$ concentration transients during alternating $CO_2$ and $H_2$ feeds.

While this fundamental study has advanced the field, their work lacks some key experimental details, such as reactor geometry (e.g., bed height, diameter, void fraction), which limits reproducibility and hinders independent validation. Additionally, the reported gas space velocity appears unusually high given the shape of the breakthrough curves, raising uncertainty about the consistency between model and experiment. Their use of a local, gradient-based least-squares method for parameter estimation, while functional, may be suboptimal for exploring complex and non-convex kinetic parameter spaces. Moreover, due to the small variation in $CO_2$ uptake with temperature in their system, the model did not fully capture the temperature-dependent trade-offs between adsorption capacity and reaction kinetics, the effects that are particularly relevant for DFMs with thermally sensitive sorbent phases.

More recently, Ono et al. [19] proposed a kinetic model based on TGA-derived conversion data for a Na-based DFM, which was later scaled to simulate reactor-scale $CO_2$ capture and methanation performance. Their approach provided mechanistic insights using an unreacted-core model and allowed analytical separation of surface reaction and diffusion contributions. However, the kinetic parameters were derived in a batch-type setup with limited gas–solid contact and strong mass transfer resistance, making their transferability to plug-flow reactors uncertain. Furthermore, system delays were only corrected in a limited post-processing step and not integrated into the kinetic fitting process.

In this work, we build upon the previous modelling approach [17], extending it to a newly developed Ca-based dual-function material (NiRu–Ca/Al) with fundamentally different sorption and reaction behaviour. Calcium-based DFMs are of particular interest due to their higher theoretical $CO_2$ storage capacity and thermal stability, making them attractive for industrially relevant temperature ranges [20], [21]. However, the kinetics and capacity profiles of such materials differ significantly from those of Na-based systems, requiring bespoke kinetic parameterisation. To address this, we present new experimental data under cyclic operation and implement a mechanistic reactor-scale model that captures the coupled dynamics of adsorption, purge, and hydrogenation. Our framework systematically incorporates the temperature dependence of $CO_2$ uptake and explicitly resolves the trade-off between storage capacity and reaction kinetics that determines the optimal operating temperature for methane production. Parameter estimation is performed using a derivative-free global search algorithm, allowing robust identification of kinetic parameters despite the model's stiffness, nonlinearities, and embedded analyser delay. Both the simulation and optimisation routines are implemented in Python and released as open source, providing a transparent and extensible platform for DFM modelling, validation, and scale-up.

In the following sections, we describe the experimental testing of a newly synthesised NiRu–Ca/Al DFM and the development of a mechanistic reactor model to simulate its cyclic performance. The kinetic model, built on first-principles transport and reaction mechanisms, is fitted to experimental data using a transparent parameter estimation framework. The results highlight key insights into stage-specific kinetics and their dependence on process parameters, followed by a discussion of modelling implications and directions for future work.

## 2. Experimental Methods

To support the development and validation of the kinetic model, a series of $CO_2$ capture and methanation experiments were conducted at different temperatures using a newly synthesised DFM composed of 15 wt% Ni, 1 wt% Ru, and 10 wt% Ca supported on alumina spheres (hereafter referred to as NiRu–Ca/Al). It is worth noting that this DFM differs slightly in shape and support type from that used in our lab's previous study [9].

This chapter outlines the material synthesis procedure, reactor setup, and the experimental conditions used to generate dynamic concentration profiles across adsorption, purge, and hydrogenation stages. These data serve as the basis for parameter estimation and model benchmarking in the subsequent sections.

### 2.1. DFM Synthesis

The DFM sample was prepared in spherical shape using a 1 mm diameter alumina support and was synthesised using a co-impregnation method. Precise amounts of $Ca(NO_3)_2·4H_2O$ (Sigma-Aldrich), $Ni(NO_3)_2·6H_2O$ (Acros Organics), and $Ru(NO)(NO_3)_3$ solution (1.5 w/v% Ru, Alfa Aesar) were dissolved in deionised water. Alumina spheres (1.0/160, Sasol) were added to the solution and stirred at room temperature for 30 minutes. Excess water was removed via rotary evaporation under reduced pressure (100 mbar) at 60–70 °C. The sample was then oven-dried overnight at 120 °C, followed by calcination at 500 °C for 3 hours using a heating ramp of 5 °C/min. The final material contained 15 wt% Ni, 1 wt% Ru, and 10 wt% Ca, and is henceforth referred to as 15Ni1Ru, Ca/Al.

## 2.2. Reactor Apparatus and System Configuration

The experimental runs were conducted using a custom-built, portable fixed-bed reactor system (see Figure 1). The reactor consists of a 1-inch OD and 20 cm horizontal stainless-steel tube (22.16 mm ID), externally heated by a dual-element heating tapes (BriskHeat) capable of maintaining temperatures up to 600 °C. A PID-controlled thermocouple embedded near the outside of the reactor wall regulates the heating, while an additional K-type thermocouple placed within the catalyst bed provides real-time logging.

Gas flows are regulated using calibrated rotameters (Aalborg) with check valves to prevent backflow. The inlet gases ($N_2$, $CO_2$, and $H_2$ mixtures) are directed to a gas-mixing chamber before entering the reactor. A manual valve system enables switching between bypass and reaction modes for calibration and leak testing.

Post-reactor, the gas stream is passed through a condenser unit comprising a catch pot and silica gel trap to remove water in the form of liquid, followed by a 2 μm particle filter. The particle-free gas is then analysed in real-time by an FT-IR gas analyser (Gasmet GT6000 Mobilis), enabling quantification of $CO_2$, $H_2$, $CH_4$, $CO$, and $H_2O$ at 5-second intervals. The setup operates at atmospheric pressure and includes a pressure relief valve and dual pressure gauges for safety. Quartz wool beds were used at both ends of the catalyst bed to ensure even flow distribution and prevent particle displacement.

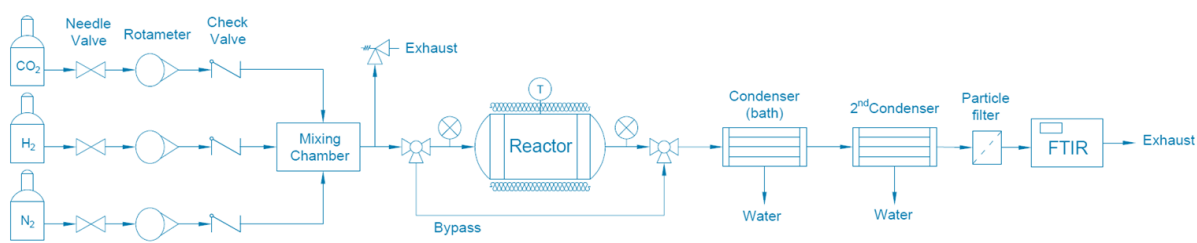

*Figure 2: Schematic of the laboratory reactor system used for $CO_2$ capture and methanation experiments for DFM*

## 2.3. Experimental Procedure for DFM Activity Testing

For the activity tests, 2 g of the DFM sample was placed in a stainless-steel fixed-bed reactor. The sample was reduced by flowing 10% $H_2/N_2$ (202 mL/min, space velocity (SV) of 52.33 min$^{-1}$) while ramping the temperature to 380 °C at 15 °C/min using external heating tapes. To remove any adventitious $CO_2$ adsorbed from ambient air, the sample was held under 10% $H_2/N_2$ at 380 °C for 20 minutes to decompose the carbonates to CO and $CH_4$ [9], [22]. After this pre-treatment, the system was flushed with $N_2$ (181 mL/min, SV of 181.41 min$^{-1}$) for 20 minutes to purge residual $H_2$, completing the reduction and conditioning cycle.

Figure 3 shows the recorded outlet concentrations of $CO_2$ and $CH_4$ during a typical cycle at 380 °C. Each test cycle includes three stages: adsorption, purge, and hydrogenation. During the adsorption phase, a 12.2% $CO_2/N_2$ stream (202 mL/min) was introduced for 20 minutes. This was followed by a 15-minute purge with $N_2$, and then a 20-minute hydrogenation step using 10% $H_2/N_2$ (202 mL/min).

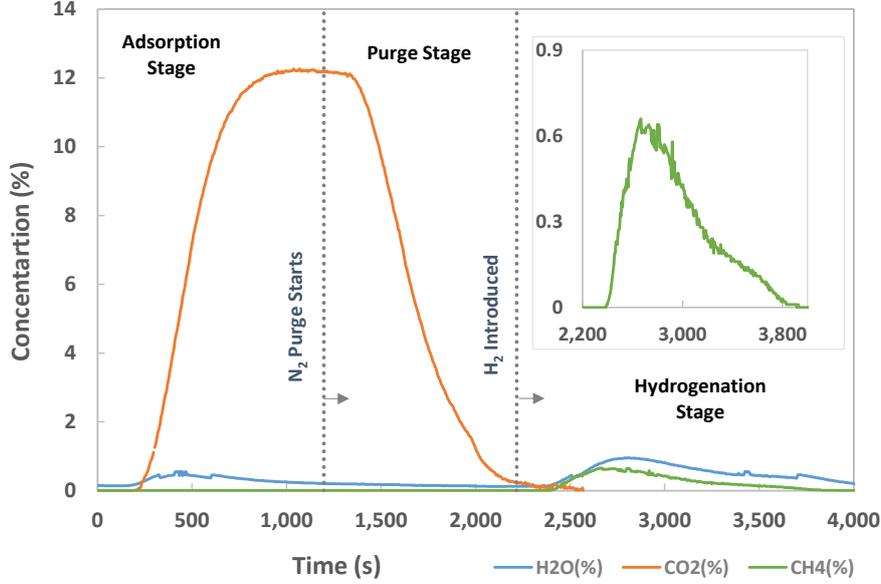

*Figure 3: Measured outlet concentrations of CO₂, H₂O, and CH₄ across the adsorption, purge, and hydrogenation stages during a single experimental cycle at 380°C. The CH₄ profile is magnified for improved visualisation due to its lower relative concentration.*

The experiment was repeated at two additional temperatures (220 °C and 300 °C). While the nominal concentrations of $CO_2$ and $H_2$ were intended to remain constant, slight variations occurred due to the challenges of manual flow regulation in the setup. Furthermore, the FT-IR analyser is not capable of quantifying hydrogen, and, according to the manufacturer's specifications, it must be protected from condensed water. To address this, the setup includes a condensation unit followed by a desiccant trap positioned upstream of the analyser, ensuring that moisture is effectively removed and measurement accuracy is maintained.

Trace amounts of carbon monoxide (CO) were detected during both the adsorption and hydrogenation stages. During adsorption, CO formation was measured at 21.3 µmol g⁻¹$_{DFM}$ and 10.5 µmol g⁻¹$_{DFM}$ at 380 °C and 300 °C, respectively. In the hydrogenation stage, the corresponding values were 15.0 µmol g⁻¹$_{DFM}$ and 1.5 µmol g⁻¹$_{DFM}$. At 220 °C, CO formation was below the detection limit for both stages, indicating negligible contribution under these conditions. Given that the CO levels were consistently low relative to the major components ($CO_2$, $CH_4$, and $H_2O$), they were not incorporated into the kinetic model or simulation framework.

Although water traps were installed in the gas line for precautionary reasons, condensation of water vapour did not occur under the operating conditions of this study. During hydrogenation, the water concentration in the effluent stream remained below 1.0 vol%, which is well under the saturation threshold at ambient temperature and pressure (>~2%). Consequently, water remained in the gas phase throughout, allowing direct quantification by FT-IR without loss due to condensation.

To quantify the $CO_2$ adsorption and methanation performance of the material across cycles, the following expressions were used:

$$\Omega_{CO_2} = \frac{1}{W} \int_0^{t_{ads}} F_{CO_2}^{in} - F_{CO_2}^{out}(t) \, dt \qquad (1)$$

$$Y_{CH_4} = \frac{1}{W} \int_{t_{prg}}^{t_{hyd}} F_{CH_4}^{out}(t) \, dt \qquad (2)$$

where $\Omega_{CO_2}$ is the maximum CO₂ adsorption capacity, $F_i^{in}$ and $F_i^{out}$ are the total flow rate for gas component $i$ entering and exiting the reactor respectively, and $Y_{CH_4}$ is the methane yield per gram of DFM sample, with $W$ being the sample mass.

To ensure accurate determination of the adsorption capacity at each temperature, the duration of the CO₂ storage (adsorption) stage was extended until the catalyst reached saturation. Similarly, the hydrogenation period was prolonged to allow complete regeneration of the sorbent. These extended durations ensured that the full capacity and reaction potential of the material were captured under each condition. A summary of the operational parameters is provided in Table 1.

*Table 1: Experimental conditions and design parameters*

| Parameter | Description | Value | Unit |
|---|---|---|---|
| $F_{total}^{in}$ | Total gas inlet flowrate at ambient temperature | 202[1] | mL/min |
| $D_i$ | Inside diameter of the reactor tube | 2.216 | cm |
| $W_{DFM}$ | Total weight of DFM granules inside the tube | 2.0 | g |
| $L$ | Reactor bed length | 1.0 | cm |
| $D$ | Diffusion coefficient | 0.16 | cm²/s |
| $\varepsilon$ | Bed voidage | 0.35 | - |
| $R$ | Universal gas constant | 8.314 | J.K⁻¹.mol⁻¹ |
| $P$ | Total Pressure (for all the three stages) | 1.0 | atm |
| $T_{amb}$ | Ambient Temperature | 293 | K |
| $T_r$ | Reactor temperature | Var.[2] | K |
| $C_{CO2,ads}$ | Concentration of CO₂ (during adsorption stage) | 12.2 | % (Vol.) |
| $C_{H2,hyd}$ | Concentration of H₂ (during methanation stage) | 10.0 | % (Vol.) |
| $N_x$ | Number of spatial steps for simulation | 5 | - |
| $N_t$ | Number of time steps for simulation | Var.[3] | - |
| $t_{ads}$ | Duration of the adsorption stage | > 1200 | s |
| $t_{prg}$ | Duration of the purge stage | > 900 | s |
| $t_{hyd}$ | Duration of the hydrogenation stage | > 1800 | s |

**Notes:**
1. Flowrate (inside the reactor) are calculated acc. to the relevant temperatures.
2. Three temperature sets of 653 K, 573 K, 493 K.
3. 3000, 2000 and 5000 for adsorption, purge and hydrogenation respectively.

It must be noted that although the reactor tube can potentially accommodate up to 20 cm of the catalytic bed length, the catalyst bed was deliberately restricted to 2.0 g of DFM over a 1.0 cm segment to approximate differential reactor conditions. This setup was selected to enable the accurate estimation of intrinsic kinetic parameters by minimising the influence of heat and mass transfer effects. The short bed length ensures near-isothermal operation, negligible pressure drop, and a low residence time, resulting in limited conversion and minimal axial concentration gradients. These conditions ensure that the measured outlet profiles reflect reaction kinetics under quasi-uniform reactant concentrations. While larger beds are indeed needed for spatiotemporal profiling or scale-up studies, the present configuration is appropriate for isolating kinetic behaviour a standard practice in DFM kinetics research which is in line with other prior parameter estimation studies [17], [19].

## 3. Computational Implementation

In this section, a one-dimensional (1D) finite difference model is developed to simulate the concentration changes in both gas and solid phases across the reactor length. The model aims to simulate and replicate the concentration profiles observed during the adsorption, purge and hydrogenation stages, such as those shown in Figure 3.

To develop this model, we incorporate the underlying mass transfer modelling principles in both gaseous and solid phase states to form a system of partial differential equations (PDEs). Based on our understandings from the mechanism of $CO_2$ capture and utilisation in DFM, we define rate expressions and integrate them into the first-principles model. The rate expressions' parameters (kinetic parameters) are identified through fitting the model on to their relevant concentration profile curves for each component in each process stage (i.e., adsorption, purge and hydrogenation). These rate expressions with their parameters fitted into the laboratory data form a set of bespoke equations that, in total, can represent the bespoke kinetic behaviour of the particular DFM formulation under various adsorption, purge and hydrogenation stages.

### 3.1. Model formulations

#### 3.1.1. Component Mass Balance

To simulate the spatiotemporal behaviour of the cyclic adsorption and reaction stages, a 1D fixed-bed reactor model was developed. The model accounts for mass balances of both gas-phase and adsorbed-phase species and forms the foundation for the kinetic parameter estimation framework. The system is represented by a set of coupled PDEs, describing transient axial dispersion, convection, and reaction phenomena in the gas phase, along with the dynamic surface coverage changes in the adsorbed phase.

The mass balance for each gas-phase species $i$ in a one-dimensional, axially dispersed plug flow reactor is described by:

$$\frac{\partial C_i}{\partial t} = -\frac{u}{\varepsilon}\frac{\partial C_i}{\partial x} + \frac{D}{\varepsilon}\frac{\partial^2 C_i}{\partial x^2} + \frac{\rho}{\varepsilon}\sum_{k=1}^{n} r_k\, \gamma_{i,k} \qquad (3)$$

where $\varepsilon$ is the reactor bed void fraction, $C_i$ is the concentration of component $i$ in the gas phase (mmol cm$^{-3}$), $u$ is the superficial gas velocity (cm s$^{-1}$), $D$ is the axial dispersion coefficient (cm$^2$ s$^{-1}$), $\rho$ is the bulk bed density (cm$^3$ g$^{-1}$) and $x$ is the distance along the reactor axis (cm).

In this equation, the term $\sum_{k=1}^{n} r_k\, \gamma_{i,k}$ represent the net rate of production or consumption of each species $i$ in the gas phase. $r_k$ is the intrinsic rate of reaction $k$ (mmol g$^{-1}$ s$^{-1}$) and $\gamma_{i,k}$ is the corresponding stoichiometric coefficient of component $i$ in reaction $k$.

The adsorbed-phase balance for surface species $j$ is described by Equation (4):

$$\frac{\partial \theta_j}{\partial t} = \frac{1}{\Omega_j} \sum_{k=1}^{n} r_k \, \gamma_{i,k} \tag{4}$$

where, $\theta_j$ is the fractional surface coverage of species $j$, and $\Omega_j$ is the adsorption capacity of the DFM (mmol g$^{-1}$).

The reactor model assumes axially dispersed plug flow with no radial gradients and negligible pressure drop, consistent with typical fixed-bed reactor modelling approaches for catalytic systems [23].

A critical consideration in kinetic modelling is the influence of heat and mass transport limitations, which can obscure the measurement of intrinsic reaction rates. In this study, the reactor configuration and operating conditions were deliberately chosen to promote kinetic control. The use of a short, differential bed helps minimise axial temperature gradients, while the combination of small catalyst particles (1 mm) and high interstitial gas velocity is intended to reduce both external and internal mass transfer resistances. This design ensures that the measured outlet profiles primarily reflect surface kinetics, enabling accurate and transferable parameter estimation.

*Boundary and Initial Conditions*

At the inlet ($x = 0$), a Danckwerts-type boundary condition was applied to account for the effect of axial dispersion while ensuring mass conservation at the inlet. The Danckwerts condition balances the convective flux with the diffusive flux [24] and is given by the following equation:

$$uC_i\,(x=0,t) - D\frac{\partial C_i}{\partial x}(x=0,t) = uC_i^{in}(t) \tag{5}$$

where $C_i^{in}(t)$ is the inlet concentration of species $i$ at time $t$.

At the outlet ($x = L$), a zero-gradient (Neumann) boundary condition was assumed, which implies no concentration gradient at the reactor exit, corresponding to fully developed flow:

$$\frac{\partial C_i}{\partial x}(x=L,t) = 0 \tag{6}$$

The initial conditions for both gas-phase concentrations and surface coverages were set to reflect the conditions at the end of the preceding experimental stage or pre-treatment step. Specifically, the final states of both the gas phase and adsorbed-phase species from each stage were used as the initial conditions for the subsequent stage. This approach ensures continuity across the adsorption, purge, and hydrogenation steps, allowing the simulation to capture the cyclic and transient nature of the process under conditions that realistically mimic the experimental sequence.

### 3.1.2. Reaction Mechanism and rate expressions

The reaction mechanism for $CO_2$ capture and in situ hydrogenation on the NiRu-Ca/Al DFM is hypothesised based on operando DRIFTS-MS insights and cyclic performance data [25], revealing distinct mechanisms across the adsorption, purge, and hydrogenation stages. Below, each stage's mechanism is detailed, mirroring the structured approach of Bermejo-López et al. [17], adapted to the NiRu-Ca/Al system.

*Adsorption Stage*

The adsorption process follows a dual-pathway, involving CaO carbonation and NiRu-facilitated $CO_2$ activation. The overall adsorption rate is governed by CaO reactivity, hydroxylation effects in the presence of $H_2O$, and transient $CO_2$ interactions with NiRu active sites.

$CO_2$ uptake primarily occurs via carbonation of CaO, forming stable calcium carbonate:

$$CaO_{(s)} + CO_{2(g)} \rightleftarrows CaCO_{3(s)} \tag{R1}$$

This reaction is rapid and highly favoured due to the strong $CO_2$ affinity of CaO, ensuring stable capture capacities. Operando DRIFTS spectra [25] provide insight into the adsorbed $CO_2$ species, confirming the presence of two distinct carbonate species. Ionic carbonates are strongly bound and require high temperatures for desorption, while monodentate carbonates are more weakly bound and contribute more significantly to $CO_2$ release in the purge step and spillover.

In the presence of $H_2O$, an alternative hydroxylation-carbonation pathway is observed:

$$CaO_{(s)} + H_2O_{(g)} \rightleftarrows Ca(OH)_{2(s)} \tag{R2}$$

$$Ca(OH)_{2(s)} + CO_{2(g)} \rightleftarrows CaCO_{3(s)} + H_2O_{(g)} \tag{R3}$$

This hydroxylation pathway is supported by $CO_2$ temperature-programmed desorption (TPD) data [25], which shows $H_2O$ desorption peaks absent under dry conditions. The presence of water promotes the formation of hydroxylated CaO sites (represented as $Ca(OH)_2$), which carbonate more readily, although their contribution remains minor compared to direct CaO carbonation. These hydroxylated sites originate either from the adsorption of atmospheric moisture or residual water retained from the previous hydrogenation (reduction) stage.

With the insights gained from the above mechanisms, the rate of $CO_2$ removal from the gas phase due to adsorption is expressed as:

$$(r_{CO_2})_{ads} = -k_1 C_{CO_2}(1 - \theta_{CO_2} - \theta_{H_2O}) - k_2 C_{CO_2} \theta_{H_2O} \tag{7}$$

which mainly accounts for CaO carbonation and its rate has proportionality to $CO_2$ concentration. The first term represents adsorption of $CO_2$ onto free CaO sites according to (R1), while the second term accounts for the replacement of the $H_2O$ molecules with competing $CO_2$ and the decomposition of the hydroxylated sites and formation of the carbonates as described in (R3).

The rate of $H_2O$ adsorption and desorption is given by:

$$(r_{H_2O})_{ads} = k_2 C_{CO_2} \theta_{H_2O} - k_3 C_{H_2O}(1 - \theta_{CO_2} - \theta_{H_2O}) \tag{8}$$

where the first term represents the release of $H_2O$ due to its replacement with $CO_2$ (the second term in Equation (8)), and the second term represents reaction (R2) where $H_2O$ is being adsorbed on CaO sites. As $CO_2$ adsorption is kinetically favoured over $H_2O$ [26], [27], the rate constant of the second term ($k_3$) is expected to be considerably lower than $k_1$.

In contrast to the model proposed for Na-based DFMs, our mechanism for the CaO-based system does not account for bicarbonate formation via $H_2O$ and $CO_2$ co-adsorption. The formation of stable, solid-phase calcium bicarbonate is not a recognised reaction pathway on CaO surfaces at the operating temperatures used in this study. Furthermore, our simpler mechanism, which attributes the transient $H_2O$ signal to the displacement of surface hydroxyls by incoming $CO_2$, sufficiently captures the experimental $H_2O$ profile without requiring additional parameters that would risk overfitting the model.

For the adsorption stage, the kinetic expressions were developed under the assumption that adsorption rates are largely insensitive to temperature. This is supported by multiple studies on high-temperature $CO_2$ chemisorption over CaO-based and alkali-modified oxides, which consistently report low apparent activation energies and fast surface-controlled kinetics [28], [29], [30]. These findings suggest that temperature has a limited effect on adsorption rates, especially in the operating temperature range of DFM processes however, it significantly influences the adsorption capacity ($\Omega_j$) of the material. In this context, the adsorption capacity is considered as a function of both gas-phase concentration and temperature. At a given temperature, its dependence on gas-phase concentration is described using a Langmuir isotherm expression (Equation (9)), where the adsorption equilibrium constant and maximum capacity are inherently temperature-dependent [24]. This formulation allows decoupling the kinetic rate terms from the adsorption capacity term, while still accounting for the exothermic nature of $CO_2$ chemisorption on basic metal oxides.

$$\Omega_j = \frac{q_{e,j}(T)\, K_{L,j}\, C_j}{1 + K_{L,j}\, C_j} \tag{9}$$

where $K_{L,j}$ is the Langmuir affinity constant, and $q_{e,j}(T)$ is the maximum adsorption capacity at equilibrium for that temperature.

To capture the temperature dependence of $q_{e,j}(T)$ the capacity must be characterised experimentally over a range of temperatures, and a material-specific polynomial fit can be developed. For moderate to high $CO_2$ concentrations (typically above 6-8 vol%), the effect of concentration on capacity becomes negligible [12], [30] and $\Omega_j$ can be approximated by $q_{e,j}(T)$, enabling a simplified pressure-independent capacity model for these conditions.

*Purge Stage*

Following $CO_2$ adsorption, a purge step is introduced to remove weakly bound $CO_2$ and residual gas-phase species, preparing the surface for the subsequent hydrogenation stage. During this stage, the flow of $CO_2$ is stopped, and an inert gas (e.g., Ar or $N_2$) is introduced to flush out unreacted $CO_2$ and $H_2O$ from the system. This step ensures that only strongly bound carbonates remain on the CaO surface while removing excess gas phase reactants that could interfere with the methanation step.

During the purge, some desorption of $CO_2$ occurs due to a decrease in $CO_2$ partial pressure, leading to the following reaction:

$$CaCO_{3(s)} \rightleftarrows CaO_{(s)} + CO_{2(g)} \tag{R4}$$

This reaction represents the decomposition of unstable carbonate species that may have formed at lower adsorption temperatures. As indicated in the adsorption stage, monodentate carbonates desorb more readily under purging conditions, contributing to the initial $CO_2$ breakthrough, while ionic carbonates require higher temperatures or prolonged purging to fully decompose. The rate of $CO_2$ desorption during the purge stage follows a Temkin-type isotherm model [31]:

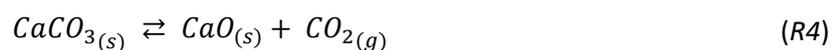

$$(r_{CO_2})_{prg} = k_4 \, exp\left(-\frac{E_4}{RT}\right) exp(1 - \alpha\theta_{CO_2})\theta_{CO_2} \qquad (10)$$

where $\alpha$ accounts for the adsorption strength of $CO_2$ onto the storage sites. As the purge proceeds, $CO_2$ coverage decreases, reducing the desorption rate over time.

Additionally, weakly adsorbed $H_2O$ desorbs from the CaO surface, as a result of the decomposition of the surface-bound hydroxyl groups which is the reverse path of reaction (*R2*).

This reaction is particularly relevant when a considerable amount of water is present during the adsorption stage (such as the steam-assisted carbonation). The rate of $H_2O$ desorption during the purge stage is similar to that of adsorption detailed in Equation (8).

For the purge stage, in contrast to the adsorption stage, the desorption rate of adsorbed $CO_2$ and $H_2O$ species is temperature-dependent, as this process involves energy-demanding bond cleavage and diffusion steps. In cases such as the purge and hydrogenation stages, where kinetics are significantly influenced by temperature, the rate constants are modelled as temperature-dependent using Arrhenius-type expressions. This allows the kinetic model to be extended or adapted for experiments conducted at different temperatures.

*Hydrogenation Stage*

Following the purge stage, hydrogen is introduced into the reactor to initiate the methanation of the previously adsorbed $CO_2$. Methanation relies on the availability of $H_2$ and the presence of active metal (Ni/Ru) sites to catalyse the conversion of surface carbonates into $CH_4$. Hydrogenation proceeds via a sequential reaction network, involving carbonate decomposition, $CO_2$ release and spillover to NiRu catalytic sites, and subsequent hydrogenation over these active sites [25].

The overall reaction for $CO_2$ hydrogenation is represented by the Sabatier reaction:

$$CO_2 + 4H_2 \rightleftarrows CH_4 + 2H_2O \qquad (R5)$$

During this stage, the $CO_2$ stored as carbonate species is first released thermally or chemically (via $H_2$ spillover) through the decomposition of $CaCO_3$, similar to the reaction specified in (*R4*).

The released $CO_2$ then undergoes hydrogenation at nearby NiRu active sites. The spillover of hydrogen from metal sites onto carbonate phases accelerates this decomposition, particularly for strongly bound carbonates such as ionic species. The $CO_2$ produced from this decomposition reacts rapidly with $H_2$ via the Sabatier reaction (*R5*), forming $CH_4$ and $H_2O$.

The rate of $CH_4$ formation is described by a potential kinetic model adapted from Lunde and Kester [32]:

$$(r_{CH_4})_{hyd} = k_5 \, exp\left(-\frac{E_5}{RT}\right)\left(P_{CO_2}^n P_{H_2}^{4n} - \frac{P_{CH_4}^n P_{H_2O}^{2n}}{K_{eq}^n}\right) \qquad (11)$$

where $k_5$ and $E_5$ are the rate constant and the apparent activation energy respectively. $P_i$ is the partial pressure of component $i$ in gas phase, $K_{eq}$ is the equilibrium constant for the Sabatier reaction, and $n$ is an empirical exponent accounting for inhibition or deviation from ideal kinetics.

The overall formation rate of gas-phase $CO_2$ is affected by both carbonate decomposition and $CO_2$ consumption in methanation, expressed as:

$$(r_{CO_2})_{hyd} = k_6 \, exp\left(-\frac{E_6}{RT}\right) \theta_{CO_2} C_{H_2} - (r_{CH_4})_{hydrogenation} \qquad (12)$$

This reflects the balance between CO₂ release from surface carbonates and its subsequent hydrogenation.

The production and removal rate of H₂O during this stage can be expressed via the following correlation:

$$(r_{H_2O})_{hyd} = 2(r_{CH_4})_{hydrogenation} - k_7 \, exp\left(-\frac{E_7}{RT}\right) C_{H_2O}(1 - \theta_{CO_2} - \theta_{H_2O}) + k_8 \theta_{H_2O} \qquad (13)$$

The second term accounts for H₂O adsorption onto free surface sites (e.g., CaO), and the third term represents water release from surface hydroxyl groups, effectively completing the water cycle from earlier adsorption.

The hydrogenation stage is essential for regenerating the CaO surface, allowing the cyclic operation of CO₂ adsorption and methanation to continue efficiently. The kinetics of this stage are influenced by the degree of catalyst reduction, temperature, and the nature of the adsorbed CO₂ species from the previous stages.

### 3.2. Kinetic Parameters Estimation

To enable estimation of the kinetic parameters governing the CO₂ adsorption and hydrogenation processes on NiRu-Ca/Al DFM, it is necessary to simulate the full dynamic response of the reactor system. The mass balances for the gas-phase and adsorbed species result in a system of PDEs that are non-linear, stiff, and spatially distributed. These equations are formed by integrating the rate expressions derived for each process stage (adsorption, purge, hydrogenation) into the component balances.

Due to the coupling between convection, dispersion, and reaction, and the presence of nonlinear source terms, analytical solutions are not feasible. Therefore, a numerical approach was adopted to simulate the evolution of gas and surface species concentrations within the reactor. The simulated outlet profiles, corresponding to the final node in a discretised 1D plug flow reactor, are then matched to experimental data through an optimisation procedure to extract the best-fit kinetic parameters.

The implementation of this numerical framework using open-source Python libraries (e.g., NumPy, SciPy) facilitates flexible model development and reproducibility. This section first introduces the modelling and numerical solution framework, followed by a description of the system response correction and, finally, the parameter estimation procedure based on fitting simulated outlet data to experimental time-resolved measurements.

#### 3.2.1. Finite Difference Method and Implicit Numerical Solver

To solve the spatiotemporal dynamics of the adsorption and hydrogenation process in the fixed-bed reactor, the system of PDEs governing gas-phase and solid-phase mass balances was discretised in space and time. Given the complexity and stiffness of the coupled kinetics, the finite difference method (FDM) was selected over other approaches, such as the finite volume method (FVM), due to its simplicity and efficiency on structured 1-D geometries, particularly for problems driven by source terms and reaction kinetics. In contrast, the FVM emphasises local conservation and is more suitable for flux-driven problems and unstructured grids. Theoretical and practical comparisons highlight that FDM offers direct handling of derivative approximations with rigorously characterised stability and convergence properties, while FVM provides enhanced conservation properties at the cost of added implementation complexity [33], [34].

The 1D column was discretised into $N$ equal spatial elements of length $\Delta x$, forming a grid along the reactor length $x \in [0, L]$. Applying FDM principles, each spatial node represents a control point where the balance equations are evaluated. The governing equations are as detailed in Section 3.1.

Spatial derivatives were discretised using second-order central difference for diffusion and first-order backward difference for convection. For a node $j$, the discretised expressions are:

$$\left(\frac{\partial c_i}{\partial x}\right)_j \approx \frac{c_i^j - c_i^{j-1}}{\Delta x}, \quad \left(\frac{\partial^2 c_i}{\partial x^2}\right)_j \approx \frac{c_i^{j+1} - 2c_i^j + c_i^{j-1}}{(\Delta x)^2} \tag{14}$$

To solve the resulting system of ordinary differential equations (ODEs) in time, an implicit Backward Euler method was employed. This method is unconditionally stable and well-suited for stiff problems [35], [36]. The generic time discretisation is:

$$\frac{y^{n+1} - y^n}{\Delta t} = f(y^{n+1}) \tag{15}$$

At each time step $n$, the nonlinear system defined by equation (15) was solved using the *fsolve* function from the *scipy.optimize* module in Python. This function employs a Newton-Raphson-based root-finding algorithm that enables the simultaneous solution of coupled, nonlinear algebraic equations without requiring an analytical Jacobian. All discretised mass and surface balance equations were combined into a residual function, which *fsolve* used to compute the next state vector $y^{n+1}$.

This fully implicit approach was chosen due to the stiffness introduced by strongly nonlinear reaction kinetics, particularly in the hydrogenation stage. Explicit time-stepping schemes such as forward Euler were avoided because they are only conditionally stable and would require prohibitively small time steps for convergence. The implicit method, despite its computational cost per step, enables significantly larger time steps while maintaining numerical robustness and stability.

### 3.2.2. Numerical stabilisation near equilibrium

Simulation of the hydrogenation stage becomes particularly challenging as the system approaches thermodynamic equilibrium, where the driving force for reaction is minimal. According to Equation (11), the thermodynamic driving force is formulated using the partial pressures of multiple components, each raised individually to the power of $n$ to reflect empirically observed kinetic behaviour. However, when the system approaches equilibrium and the driving force becomes very small, the use of a fractional exponent ($n < 1$) can lead to numerical instabilities. This arises because near-zero values raised to small powers can amplify rounding errors or produce erratic gradients.

To address this issue, a regularisation strategy was implemented. For small values of the driving force, the power-law term was replaced with a second-order polynomial approximation that is smooth and differentiable near zero. This approach maintains the physical consistency of the model while preventing spurious oscillations in the computed reaction rate under low concentrations of $CO_2$, $CH_4$, or $H_2O$ near equilibrium.

### 3.2.3. System delays and response model

In our experimental apparatus (detailed in Section 2), the long connecting tubing, presence of two condensers, and various fittings introduce a degree of back-mixing, dispersion, and dynamic delay.

Additionally, the gas analyser itself contributes to the total delay due to internal sampling, averaging, and sensor response times. These combined effects significantly alter the measured signal compared to the reactor outlet concentration profile and necessitate the application of a dynamic correction model.

To account for these effects, we applied a downstream smoothing model that captures the cumulative influence of mixing, delay, and inertia in the detection system. Specifically, we used a second-order dynamic response model, a standard approach in process control, to approximate the behaviour of the system downstream of the reactor. This model captures both the time lag and gradual rise or decay of the analyser response, especially in systems exhibiting mild back-mixing and delayed equilibration.

The output response $C_{measured}(t)$ of a second-order linear system can be modeled by the following differential equation:

$$\tau^2 \frac{d^2 C_{measured}}{dt^2} + 2\zeta\tau \frac{dC_{measured}}{dt} + C_{measured} = C_{model}(t) \qquad (16)$$

where $C_{model}(t)$ is the simulated reactor outlet concentration (ideal plug flow output), and $C_{measured}$ is the delayed concentration signal observed by the analyser. Parameters $\tau$ and $\zeta$ are the system's time constant and damping ratio respectively which define the transient behaviour of the system.

To calibrate the parameters of this response model, we used blank test data (i.e., running the experiment through the same apparatus without reactive bed and empty reactor) to isolate and characterise the delay profile independently of reaction kinetics. Separate parameter sets for the rising and falling phases ($\tau_{rise}, \zeta_{rise}$ and $\tau_{fall}, \zeta_{fall}$) were employed to better match the asymmetry typically observed in adsorption/desorption transitions. During simulation, the model selects the appropriate set of parameters at each time step depending on whether the signal is increasing or decreasing, enabling smooth and accurate reproduction of the asymmetric analyser response observed in practice, particularly during hydrogenation cycles.

Figure 4 shows the analyser's response to the step change in the inlet gas composition during the blank test. The resultant time constant and damping factors for the rise and fall transitions are specified in the Figure.

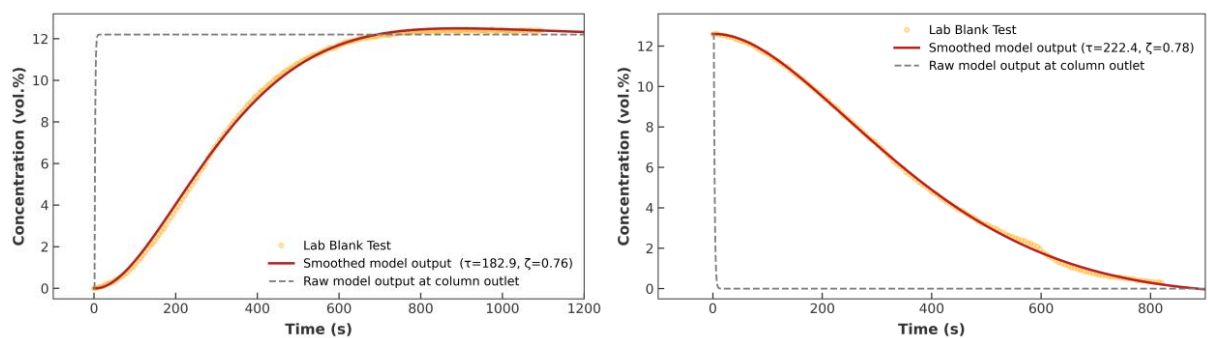

Figure 4: Second-order response to step changes in CO₂ concentration during blank tests. (a) Rising transition with fitted parameters (τ = 182.9 s, ζ = 0.76); (b) Falling transition (τ = 229.4 s, ζ = 0.80). Dashed lines show ideal model output; solid lines represent the smoothed analyser response, with the lag time being filtered out.

This second-order response model serves as a surrogate for the downstream system's inertia and provides a more realistic convolution of the modelled concentration profile. It is particularly advantageous over simple first-order lags, as it better captures the S-shaped transients observed in the experimental analyser signal without requiring detailed mechanistic modelling of the full tubing and condenser setup.

By applying this post-processing filter to the simulated outlet concentrations, the final model outputs could be directly compared with the analyser data, enabling accurate and meaningful parameter estimation in the presence of measurement system delays.

It is important to note that mild overshooting is already visible in the experimentally measured signal during the blank test, even in the absence of any reactive material. This indicates that the behaviour is not a modelling artefact, but rather a physical feature of the detection system, likely arising from internal back-mixing, gas hold-up, and the intrinsic dynamic response of the analyser.

### 3.2.4. Kinetic Parameter Estimation and Fitting Strategy

To determine the kinetic parameters associated with the model developed for the three stages of adsorption, purge, and hydrogenation, an optimisation framework was employed. This procedure integrates the numerical reactor model with the delayed response, producing a convoluted output that more accurately reflects the experimentally observed data. The resultant model output, incorporating both the reactor dynamics and measurement system delays, was then fitted to laboratory experimental data obtained for each of the three stages independently. The optimisation process aimed to identify parameter values that minimise the deviation between the model-predicted concentration profiles (specifically, the output after delay convolution) and the experimentally measured gas compositions.

Kinetic parameters were estimated using a Bayesian optimisation framework implemented via the Optuna [37] library in Python. This approach replaces conventional gradient-based techniques used in many similar parameter estimation studies [38], [39]. Unlike local, gradient-based solvers which can suffer from convergence to sub-optimal local minima, Bayesian optimisation effectively balances exploration and exploitation, enabling it to navigate complex, non-linear, and stiff objective landscapes with multiple local minima or discontinuities due to solver instabilities, thereby leading to a more robust and globally optimal fit [40]. The optimisation sought to minimise a scalarised objective function representing the weighted sum of squared residuals between model-predicted and experimental outlet concentration profiles of $CH_4$, $H_2O$, and $CO_2$.

The search over the parameter space was guided by Optuna's Tree-structured Parzen Estimator (TPE) sampler, which adaptively balances local refinement and global exploration [41]. The optimisation included integer, float, and logarithmic parameter types across activation energies, pre-exponential constants, and adsorption constants. For trials producing unstable or non-physical model outputs, a high penalty value was returned to facilitate pruning. The final best-fit parameter set was selected as the one minimising the total residual across all temperature conditions. The final cost function $J(k)$ is defined as:

$$J(k) = \sum_{i=1}^{N} w_i (C_{exp,i} - C_{model,i}(k))^2 \qquad (17)$$

where $k$ is the vector of fitted parameters, $C_{exp,i}$ and $C_{model,i}$ represent the experimental and model-predicted concentrations at time step $i$. A weighting factor $w_i$ is introduced to reduce the impact of noisy measurements, particularly during the $CO_2$ hydrogenation stage. FT-IR analysis of trace $CO_2$ often suffers from baseline fluctuations and amplified noise due to low signal strength [42]. By down-weighting these unreliable data points, the optimisation remains focused on the more accurate concentration profiles (e.g., $CH_4$, $H_2O$), ensuring that parameter estimation is not skewed by signal artefacts.

A preliminary sensitivity analysis around the kinetic parameters were performed to identify the effect of each parameter's manipulation on the outputted concentration profiles and coverage factors and also the variation bounds of the optimisation.

For the adsorption and purge the sensitivity analysis and variation bounds were guided by centring around values reported in our reference work [17].

However, for the hydrogenation step, the kinetic parameters were manipulated around the values from an independent study on the $CO_2$ methanation kinetics over 15 wt% Ni – 1 wt% Ru/$CeO_2$-$Al_2O_3$ catalyst conducted in our laboratory [43]. This study reported an activation energy of 80.9 ± 2.26 kJ mol$^{-1}$ and provided experimentally validated reaction orders for $H_2$, $CO_2$, $CH_4$, and $H_2O$. These values were derived under continuous-flow operation without adsorption storage, thereby isolating the catalytic activity of the NiRu phase.

However, in our DFM process, $CO_2$ is first captured and later hydrogenated through a spillover mechanism involving pre-adsorbed carbonate species. This temporal separation is known to modify the apparent kinetics, particularly by reducing the activation barrier due to enhanced surface mobility and local interactions [5], [44]. To account for this, we used the activation energy and pre-exponential factor from a recent study [43] within the Lunde and Kester approach-to-equilibrium formulation. Such formulation inherently accounts for equilibrium limitations and was selected for its robustness in capturing near-equilibrium behaviour. Given the mechanistic differences between the DFM hydrogenation and continuous methanation systems, the activation energy was not strictly fixed but permitted a limited degree of freedom to better match observed reaction behaviour.

The combination of experimentally validated parameters, physically interpretable constraints, and robust optimisation tools enabled accurate extraction of kinetic constants and ensured model agreement with the observed adsorption and hydrogenation dynamics.

## 4.     Results and Discussions

This section presents the results of kinetic parameter estimation based on the mathematical framework developed in Section 0. The system of PDEs governing the mass balances for components present in both gas and solid phases was solved numerically for each stage of the cyclic process. Simulated outlet concentrations (corresponding to the final axial node of the reactor domain) were corrected using a second-order analyser delay model to account for the downstream tubing delays and dynamic response of the gas detection system. These delay-adjusted model predictions were then fitted to the experimental gas composition data to extract the relevant kinetic parameters.

Parameter estimation was performed using a Bayesian optimisation approach implemented via the Optuna library in Python and the parameter estimation for each stage of the cyclic process (adsorption, purge, and hydrogenation) was conducted independently. Due to system delays (e.g., tubing dispersion and analyser response time) the kinetic features of the adsorption and specially the purge stages were affected, limiting the sensitivity of the model output to their respective kinetic parameters. Therefore, a *combined parameter estimation* strategy was employed to confidently identify the parameters of the purging process kinetics along with estimation of the parameters for the hydrogenation dynamics as the coverage factor resulting from the purging stage serves as the initial point of the hydrogenation. These two stages were coupled in the simulation, with the surface coverage remaining at the end of the purge stage serving as the input condition for the hydrogenation step.

The model was executed on a machine equipped with an 11th Gen Intel® Core™ i7-1165G7 processor (2.80 GHz, 4 cores, 8 threads). Each optimisation run was set to a maximum of 1000 trials, with the combined model taking 185 minutes to perform these. Full implementation details and access to the Python source code are provided in the Data Availability section.

### 4.1. Parameter estimation results

Figure 5 (a–c) shows the experimental outlet concentration profiles of $CH_4$, $CO_2$, and $H_2O$ over the three sequential stages of adsorption, purge, and hydrogenation, under isothermal conditions at 380 °C. The curves also display the model predictions, which were processed through a second-order delay response function to account for analyser lag and downstream dispersion. These fitted outputs enable comparison with the experimental analyser data but do not represent the true reactor outlet concentrations.

During the adsorption stage (Figure 5a), a 12.2 vol% $CO_2$ stream was introduced and maintained for 1200 seconds. The kinetic model simulates rapid capture of $CO_2$ on CaO sites via surface-controlled carbonation reactions, while a minor $H_2O$ signal is associated with the displacement of hydroxyl species and their conversion to carbonates. However, due to the smoothing effect of the delay model, the sharpness of the adsorption front and any transient peaks cannot be discerned directly from Figure 4.

Following adsorption, the purge stage commenced by switching to pure $N_2$ for 900 seconds. The model output reflects a gradual reduction in $CO_2$ signal due to desorption and slow carbonate decomposition, although the real concentration front during purge (characterised by a fast initial sweep followed by a long tail) is not directly visible in the delay-convoluted curves shown in Figure 5b.

The hydrogenation stage, initiated by introducing 10 vol% $H_2$ for 1800 seconds, shows an increase in $CH_4$ and $H_2O$ concentrations, corresponding to methanation of stored $CO_2$. As can be seen from Figure 5c the fitted curves closely track the experimental analyser data, validating the kinetic model's ability to capture the cyclic behaviour of the process. However, due to the analyser's limitations, $H_2$ was not detected experimentally and thus excluded from the fitting procedure. Additionally, while the fitted profiles suggest sustained $CH_4$ and $H_2O$ formation, the convolution masks finer kinetic details such as the onset of $CO_2$ release or the precise timing of the methanation front.

For the hydrogenation stage, the model shows a slightly weaker fit for the $CO_2$ concentration, which is attributable to the inherent challenges of measuring trace gases with infrared spectroscopy. At the low concentrations observed in this stage, the signal-to-noise ratio for $CO_2$ is reduced, leading to baseline fluctuations and lower precision in the FT-IR signal. This is a known instrumental limitation for quantitative analysis near a device's detection limits [42]. As detailed in our parameter estimation strategy, this effect was anticipated and caused by assigning a lower weighting factor to the noisier trace $CO_2$ data.

Figure 5 d to f show the temporal evolution of surface coverage for $CO_2$ and $H_2O$ ($\theta_{CO2}$ and $\theta_{H2O}$). These coverage profiles are generated directly from the kinetic model and are not subjected to the delay response function.

During adsorption (Figure 5d), $\theta_{CO2}$ increases progressively from 0 to nearly 1 as CaO sites are occupied by carbonate species. During the purge phase (Figure 5e), $\theta_{CO2}$ declines gradually, indicating the slow decomposition and desorption of unstable carbonates. Upon hydrogenation (Figure 5f), $\theta_{CO2}$ continues to decrease as hydrogen spillover facilitates methanation.

Simultaneously, water formed during the reaction is partially adsorbed on available CaO sites, reflected by a transient rise in $\theta_{H2O}$. As the process continues, some of this surface water desorbs and exits the reactor, while a small residual fraction remains adsorbed, forming the initial $\theta_{H2O}$ for the next cycle. This remaining water is gradually replaced by $CO_2$ during the subsequent adsorption stage. The minor $H_2O$ peaks observed during adsorption can thus be attributed to this surface exchange and the decomposition of hydroxylated sites in the presence of incoming $CO_2$.

Although the surface coverage profiles cannot be directly validated experimentally, they are mechanistically consistent and play a key role in bridging the three stages. The final values at the end of each stage serve as the initial conditions for the next, and their indirect impact can be observed in the quality of the fitted concentration profiles. For this reason, $\theta_{CO2}$ and $\theta_{H2O}$ are plotted without smoothing, offering insight into the true surface dynamics driving the observed behaviour.

To gain deeper insight into the spatiotemporal behaviour of the system and to validate the mechanistic assumptions underlying each stage, Figure 6 to Figure 8 present three-dimensional plots of gas-phase concentrations and surface coverages across the reactor length and over time. These figures reflect the true model outputs (i.e., unaffected by analyser delay or smoothing) and thus provide a more accurate representation of the physical concentration fronts and spatial gradients that are otherwise hidden in the delay-corrected profiles of Figure 5.

Figure 6 illustrates the adsorption stage, focusing on the first 100 seconds. In Figure 6a, the $CO_2$ surface coverage ($\theta_{CO2}$) rises almost immediately at the reactor inlet, indicating rapid carbonation of the CaO sites as the $CO_2$ feeding begins. The coverage front progressively advances along the reactor length as $CO_2$ molecules diffuse through the bed and saturate the available sites. A similar trend is observed in the gas-phase $CO_2$ profile (Figure 6b), where concentrations increase gradually downstream as the bed becomes progressively saturated.

Figure 6c shows a sharp initial decline in $\theta_{H2O}$, particularly at the reactor inlet. This reflects the displacement and decomposition of pre-existing hydroxylated CaO sites as $CO_2$ begins reacting to form carbonates. The corresponding gas-phase $H_2O$ concentration (Figure 6d) displays a distinct peak near the inlet, followed by a cumulative increase along the bed as additional water is released from deeper hydroxyl sites. As time progresses and $\theta_{H2O}$ approaches zero, the $H_2O$ concentration in the gas phase also drops, confirming the depletion of surface hydroxyls and completion of the carbonation reactions. This behaviour confirms the operation of both the direct carbonation pathway and the hydroxyl-mediated carbonation mechanism described in Reactions (*R1*)–(*R3*).

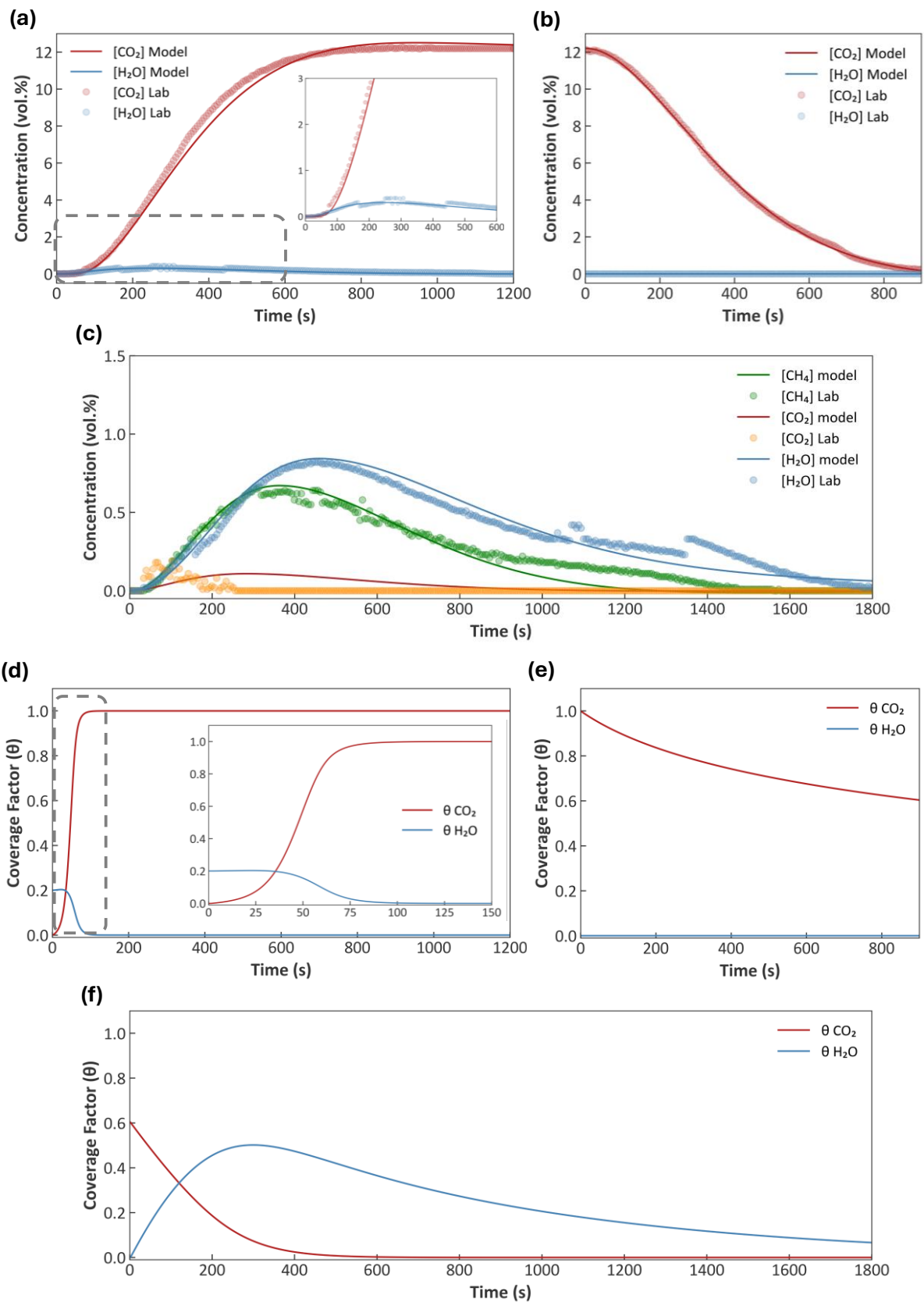

*Figure 5: Experimental and modelled outlet concentration profiles for $CO_2$, $H_2O$, and $CH_4$, along with the corresponding surface coverage dynamics during the three process stages at 380 °C: (a, d) adsorption, (b, e) purge, and (c, f) hydrogenation. Modelled concentration curves include the analyser delay and are plotted alongside experimental data, while coverage factor changes at the reactor outlet are shown without delay correction.*

Figure 7 focuses on the purge stage. Figure 7a and Figure 7c depict the full-time evolution of $\theta_{CO_2}$ and $\theta_{H_2O}$, respectively, while Figure 7b and Figure 7d show gas-phase $CO_2$ and $H_2O$ concentrations over the first 20 seconds of purging.

In Figure 7b, a sharp drop in $CO_2$ concentration is observed at the reactor inlet immediately after switching to pure nitrogen. This reflects the rapid displacement of residual gas-phase $CO_2$ by $N_2$ and corresponds to the initial breakthrough sweep. However, after this initial purge, the $CO_2$ profile flattens to near-zero levels, suggesting that any further release of $CO_2$ from the solid phase (namely, the desorption and decomposition of weakly bound carbonates) is too slow and diffuse to be detectable in the gas stream. This interpretation is supported by Figure 7(a), where $\theta_{CO_2}$ continues to decline gradually over time. Although this ongoing decomposition is kinetically relevant, its contribution to gas-phase $CO_2$ is minimal and smeared over a long period, hence remaining undetectable in the analyser-convoluted profiles.

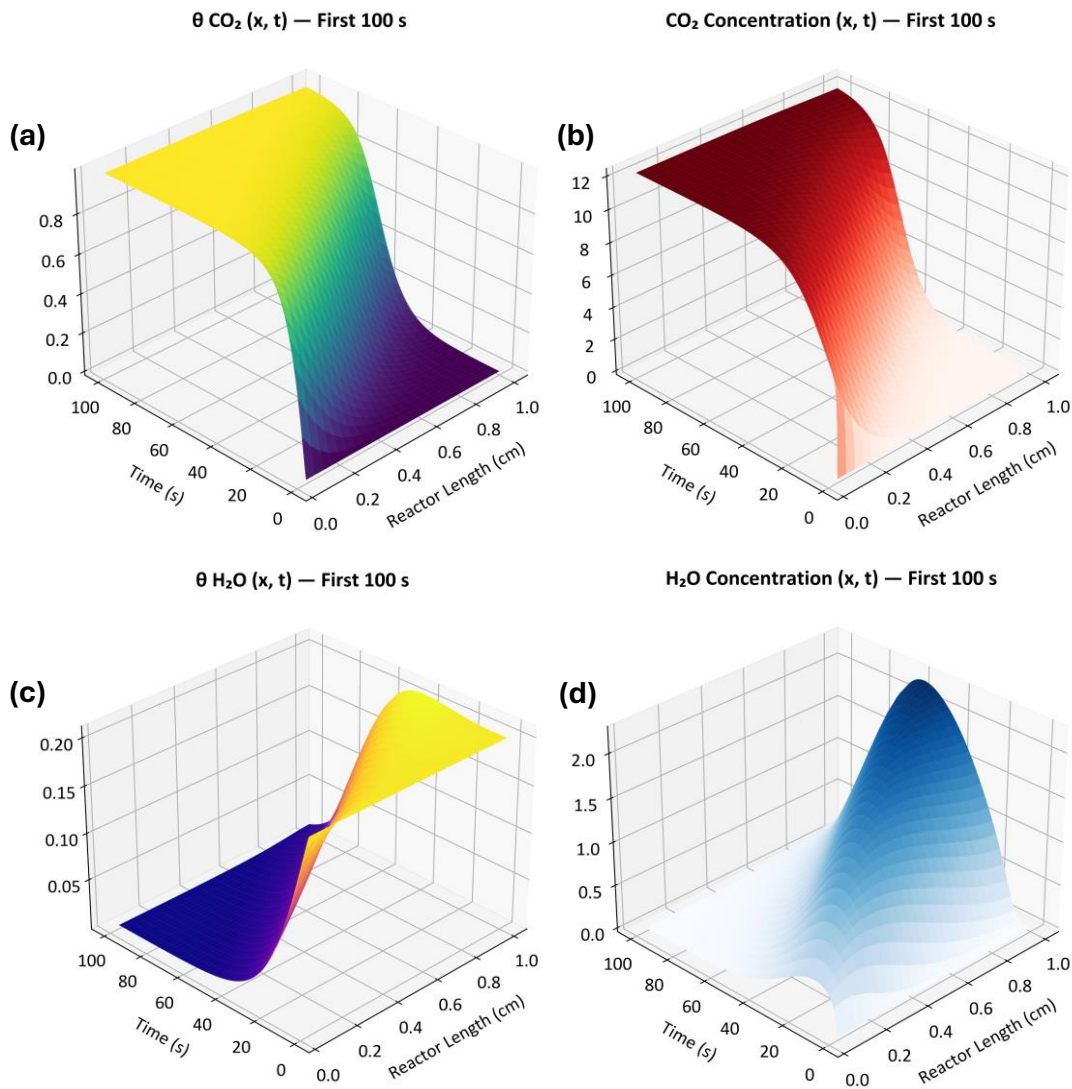

*Figure 6: Temporal changes to the concentration fronts and coverage factors across the reactor length during adsorption stage for the first 100 s of the process.*

Similarly, as shown in Figure 7(c–d), $\theta_{H2O}$ and gas-phase $H_2O$ remain effectively flat during the purge. This is consistent with the fact that most water-forming reactions had already completed during the adsorption stage, leaving no residual hydroxyl species to contribute to further $H_2O$ desorption.

Figure 8 presents the hydrogenation stage and the model output illustrates the behaviour of all major components ($CO_2$, $CH_4$, $H_2$, and $H_2O$) over time and space, as well as their corresponding surface coverages.

As shown in Figure 8d, a sharp hydrogen front enters the reactor and initiates contact with the carbonate-loaded sorbent. This immediately activates the spillover mechanism, triggering the decomposition of calcium carbonates and subsequent methanation. Figure 8e shows a corresponding decline in $\theta_{CO2}$, indicating the progressive release of $CO_2$ from the surface. However, Figure 8a shows that gas-phase $CO_2$ concentrations remain low and transient, as $CO_2$ is rapidly consumed upon formation due to the high activity of Ni/Ru catalytic sites.

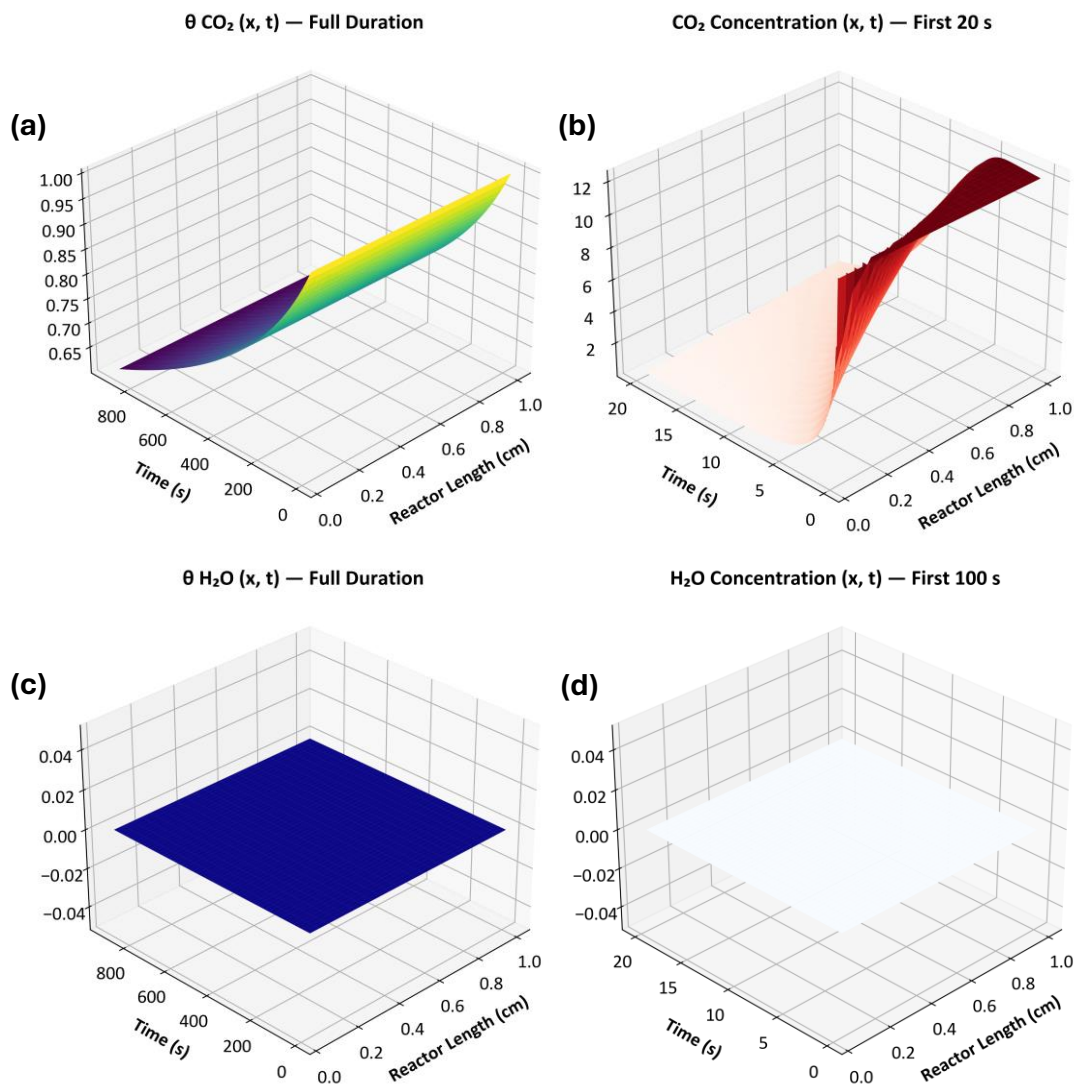

Figure 7: Temporal changes to the concentration fronts and coverage factors across the reactor length during purge stage. CO2 concentration front is depicted for the first 20 s of the process

Figure 8c displays the gradual and sustained formation of CH₄, suggesting that methanation proceeds efficiently throughout the reactor. Importantly, this continues even after gas-phase $CO_2$ becomes nearly undetectable, which highlights that carbonate decomposition, not methanation itself, is the rate-limiting step under these conditions. Figure 8b shows the corresponding increase in gas-phase $H_2O$, confirming stoichiometric water formation. In parallel, a portion of this $H_2O$ adsorbs onto CaO, as seen in the rise of $\theta_{H2O}$ in Figure 8f. Over time, $\theta_{H2O}$ stabilises, establishing the initial coverage condition for the next cycle.

## 4.2. Response time and combined parameter estimation approach

As previously discussed, the kinetics of $CO_2$ adsorption on CaO-based adsorbents are relatively fast and can be significantly masked by the physical configuration of the experimental setup. In particular, long tubing, the presence of equipment sets, and the intrinsic delay of the gas analyser introduce substantial system lag, which distorts the observed breakthrough curves. This convolution of system response with the true reactor dynamics presents a serious limitation in the accurate estimation of kinetic parameters, especially during fast transients such as adsorption and early-stage purge.

The purge phase, although initiated by a rapid sweep of trapped gases from the reactor, is followed by a gradual decline in surface coverage due to the slow decomposition of unstable carbonate species. The $CO_2$ released during this phase is relatively small in quantity and often remains below the detection threshold of the analyser due to the smoothing effects of system delay. As a result, its contribution to the outlet concentration profile is minimal and difficult to resolve, rendering the convoluted curve highly insensitive to changes in the true kinetic behaviour of the purge stage.

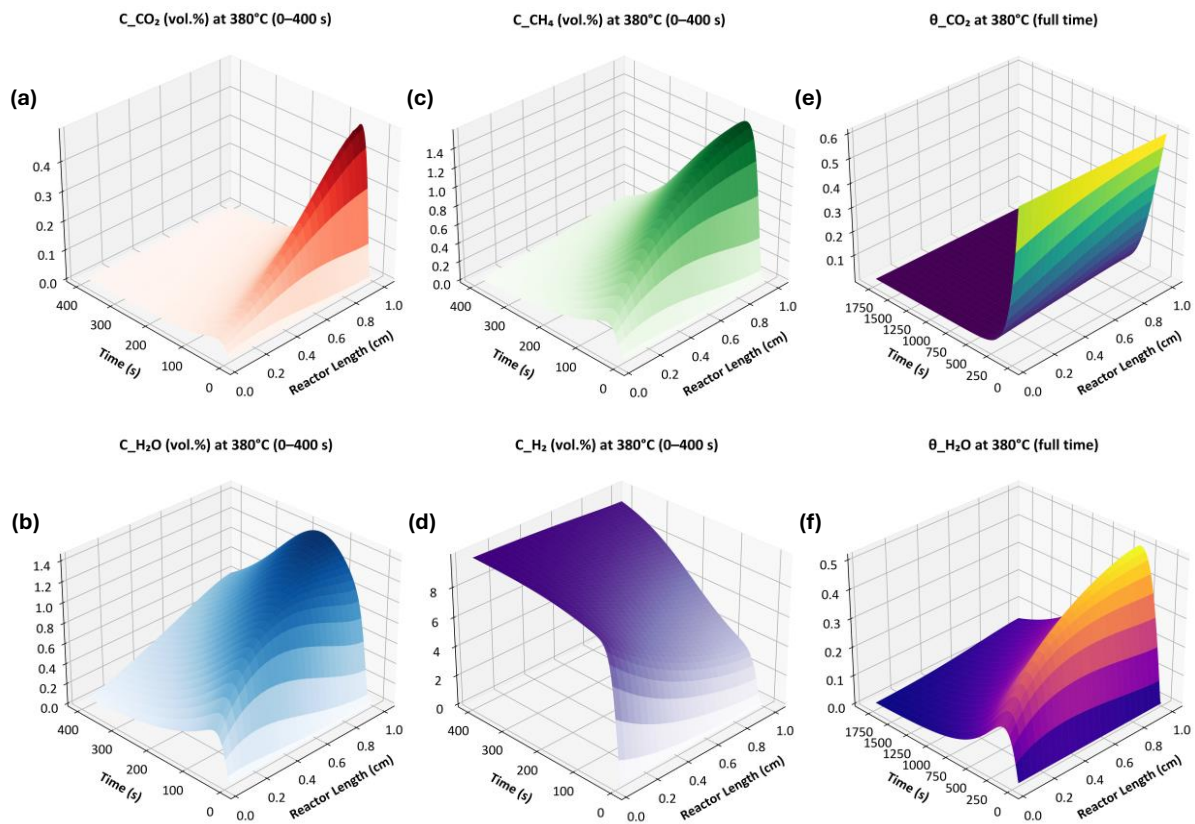

*Figure 8: Temporal changes to the concentration fronts and coverage factors across the reactor length during methanation stage for the first 400 s of the process. The coverage factors are shown for the full time length.*

This masking effect is evident when examining the fitted profiles in Figure 5. The adsorption and purge stages exhibit better agreement with the experimental data compared to the hydrogenation stage, but this alignment is somewhat misleading. For adsorption, the intrinsic kinetics are so fast that they become indistinguishable from the delay-dominated system response. Similarly, during purge, the sharp initial drop in $CO_2$ concentration is captured, but the subsequent slow desorption phase (while mechanistically significant) results in a shallow gradient that the analyser cannot reliably detect. This insensitivity limits the model's ability to extract meaningful kinetic parameters from the outlet signal, particularly for the purge process.

During parameter estimation for the adsorption stage, it was observed that the convoluted model's output remained sensitive to the adsorption rate constant $k_1$ only up to a threshold of approximately 60 $cm^3\,s^{-1}\,g^{-1}$. Below this value, increasing this parameter produced progressively sharper adsorption fronts in the simulated profile. However, beyond this threshold, further increases had negligible impact on the predicted response. This plateau in sensitivity reflects the dominance of analyser lag and dispersion, which smooth out the fast intrinsic kinetics and limit the model's temporal resolution. Once the reaction rate exceeds this resolution, it becomes effectively undetectable in the observed output.

Interestingly, this threshold value aligns with that reported in our reference study [17] indicating that the fitted value for $k_1$ lies within a physically reasonable range. To prevent overfitting or inferring artefacts beyond the system's detection capability, $k_1$ was fixed at this value. More accurate determination of the intrinsic adsorption kinetics would require future experiments with reduced system lag, such as shorter tubing and higher-frequency detection systems.

In contrast, the purge stage demonstrated even lower sensitivity to its rate-controlling parameters. Nevertheless, these parameters are critical because they determine the final $CO_2$ surface coverage, which serves as the starting condition for the hydrogenation stage. To address this, a *combined parameter estimation approach* was adopted, wherein the purge and hydrogenation stages were fitted simultaneously rather than independently. In this framework, the remaining surface coverage from the purge simulation was passed forward as the initial condition for hydrogenation in each iteration, ensuring mass continuity across stages.

This coupling strategy allowed the identification of a parameter set that not only reproduced the $CH_4$ production trends but also yielded physically meaningful $CO_2$ surface coverages at the end of purging. In doing so, the purge kinetics were indirectly validated through their influence on downstream methanation behaviour. The final fitted kinetic parameters for all three stages are summarised in Table 2.

*Table 2: The fitted values of the kinetic parameters resulting from the optimisation work.*

| Parameter | Value | Unit | Note |
|---|---|---|---|
| $k_1$ | 60 | $cm^3\,s^{-1}\,g^{-1}$ | |
| $k_2$ | 20 | $cm^3\,s^{-1}\,g^{-1}$ | |
| $k_3$ | 1.0 | $cm^3\,s^{-1}\,g^{-1}$ | |
| $k_4$ | 0.012 | $mmol\,g^{-1}\,s^{-1}$ | |
| $E_4$ | 34 | $J\,mmol^{-1}$ | |
| $k_5$ | 43.96 x $10^3$ | $mmol\,g^{-1}\,s^{-1}\,atm^{-5n}$ | |
| $E_5$ | 65 | $J\,mmol^{-1}$ | |

| Parameter | Value | Unit | Note |
|---|---|---|---|
| $k_6$ | 38 | cm$^3$ s$^{-1}$ g$^{-1}$ | |
| $E_6$ | 14 | J mmol$^{-1}$ | |
| $k_7$ | 144 | cm$^3$ s$^{-1}$ g$^{-1}$ | |
| $E_7$ | 10 | J mmol$^{-1}$ | |
| $k_8$ | 0.003 | cm$^3$ s$^{-1}$ g$^{-1}$ | |
| $\alpha$ | 0.5 | - | |
| $n$ | 0.14 | - | (1) |

Note (1): This value is fixed based on [32]

## 4.3. Effect of temperature on the process kinetics

The cyclic CO$_2$ capture, purge, and hydrogenation process was also conducted at two additional temperatures: 300 °C and 220 °C, to investigate the influence of temperature, particularly on the purge and hydrogenation stages. While temperature has a relatively minor impact on the kinetic parameters of the adsorption stage, it has a pronounced effect on both purge and hydrogenation dynamics. One key parameter affected during adsorption, however, is the maximum CO$_2$ adsorption capacity of the DFM, which directly influences the coverage factor and, consequently, the overall performance of the cycle.

Figure 9 compares the experimentally measured CO$_2$ adsorption capacities as a function of temperature for two closely related dual-function materials: the DFM synthesised in this study, and the formulation reported by Merkouri et al. [9] with similar metal and sorbent loadings, differing only in the addition of CeO$_2$ to the support. The literature data are included here as the closest available reference, with minimal compositional differences expected to affect adsorption behaviour. The figure shows that the adsorption capacity decreases significantly from 220 °C to approximately 450 °C, consistent with the exothermic nature of the carbonation reaction and the thermodynamic preference for carbonate formation at lower temperatures. This decline aligns with prior studies indicating reduced CaO activity capture efficiency at higher temperatures due to sintering, pore closure, and diminished surface area [45]. Interestingly, a modest increase in CO$_2$ uptake is observed beyond 450 °C, potentially due to a transition from kinetic to equilibrium-limited behaviour or surface restructuring effects such as partial carbonate decomposition and reformation, which have been noted at elevated temperatures in high-loading CaO systems [45]. However, further mechanistic studies would be necessary to fully confirm this phenomenon in DFM systems.

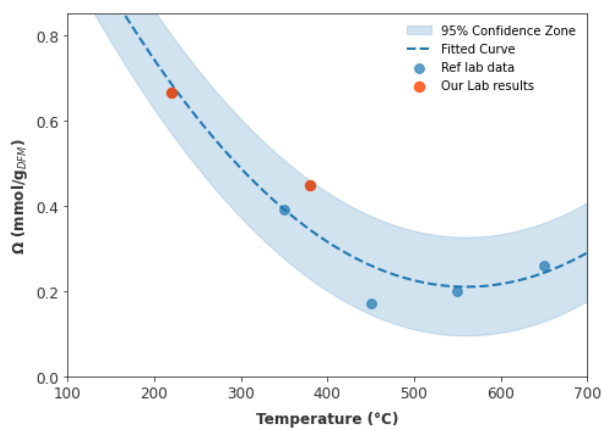

*Figure 9: Temperature dependence of $CO_2$ adsorption capacity for two closely related DFMs. The data include results from this study and from a previously reported formulation [9] with almost similar composition, except for the addition of $CeO_2$ to the support.*

Table 3 summarises the measured $CO_2$ adsorption capacities and the corresponding initial $\theta_{CO_2}$ values at the start of the hydrogenation stage (i.e., at the end of the purge phase) for the three tested temperatures. These simulations were based on the fitted kinetic parameters listed in Table 2, and the resulting outlet concentration profiles of $CH_4$, $H_2O$, and $CO_2$ are illustrated in Figure 10.

*Table 3: Changes in maximum $CO_2$ adsorption capacity and initial $CO_2$ surface coverage for three temperature sets.*

| Temperature (°C) | $\Omega_{CO_2}$ (mmol g$^{-1}$) | Initial $\theta_{CO_2}$ |
|---|---|---|
| 220 | 0.665 | 0.85 |
| 300 | 0.550 | 0.79 |
| 380 | 0.448 | 0.68 |

As shown in the simulated profiles, the model successfully predicts $CH_4$ production trends across all three temperatures. Interestingly, methane yield reaches a maximum at 300 °C, despite the expectation (based on Arrhenius correlation) that higher temperatures should enhance reaction rates. This non-monotonic trend results from the competing effects of temperature-dependent methanation kinetics and adsorption capacity. While higher temperatures accelerate methanation, they also reduce the amount of stored $CO_2$ available for conversion. This trade-off highlights the existence of an optimal temperature that balances kinetic enhancement with sorbent loading to maximise methane production.

It is important to remember that the activation energy and pre-exponential factor for the methanation rate were not fixed to values obtained from previous continuous-flow experiments on the 15 wt% Ni – 1 wt% Ru/CeO$_2$-Al$_2$O$_3$ catalyst [43]. Instead, they were treated as free parameters in the fitting process. In our continuous-flow benchmark study [43], an activation energy of 80.9 J mmol$^{-1}$ was reported. In contrast, the parameter estimation performed in this work produced a lower activation energy of 65 J mmol$^{-1}$ (approximately 20% lower), and a higher pre-exponential factor of 58.5 × 10$^3$ mmol g$^{-1}$ s$^{-1}$ atm$^{-5n}$ (about 23% higher than that of the experiment).

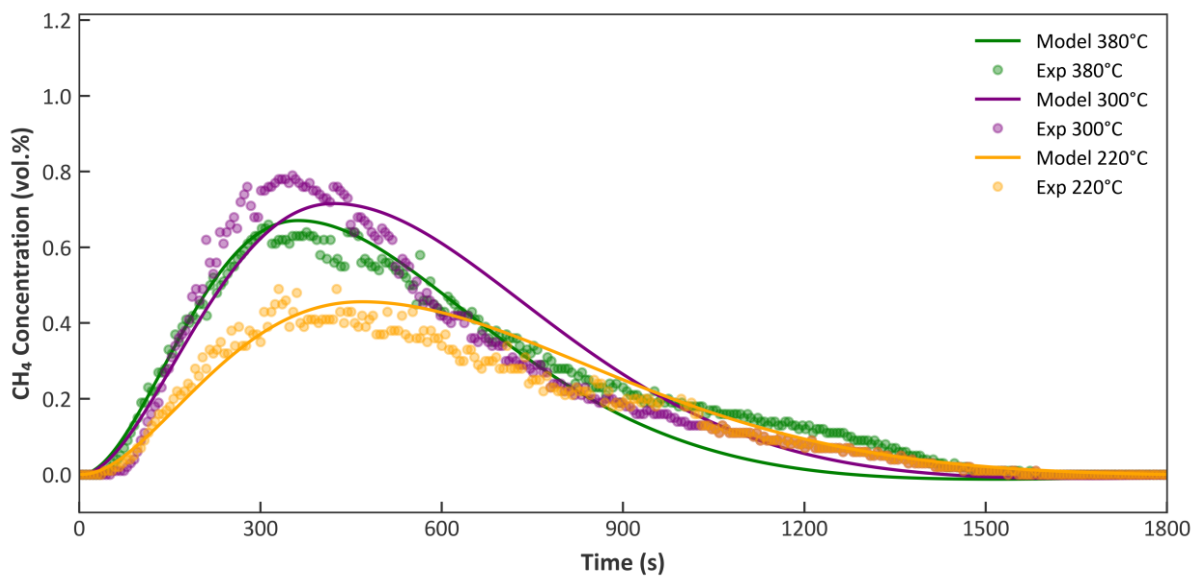

*Figure 10: Temperature effect on methanation. The methane yield from the kinetic model for three different temperatures (220 C, 300 C and 380 C) are compared with their corresponding laboratory results.*

This difference is mechanistically justifiable. The experimental benchmark was based solely on catalyst behaviour in a continuous Sabatier mode, whereas the current system operates under DFM cyclic conditions involving adsorbed carbonate decomposition and hydrogen spillover. The DFM mechanism introduces local activation effects and temporal separation of steps, which likely reduce the apparent energy barrier compared to the continuous reaction. These findings suggest that the spillover-driven decomposition of carbonates facilitates methanation under milder conditions, representing a key advantage of the DFM process over conventional Sabatier systems.

Another important observation is the consistently low rate of $CO_2$ escape during hydrogenation across all three temperature conditions. This supports the conclusion that carbonate decomposition (not methanation) is the rate-limiting step. The $CO_2$ generated during decomposition is rapidly consumed via methanation (i.e., $r_{CH_4} > r_{CO_2}$), which explains the minimal breakthrough observed in gas-phase $CO_2$ during this stage.

To reduce uncertainty in future experiments, the downstream tubing length should be minimised, and the use of auxiliary components should be avoided wherever possible to decrease system lag. Extending the reactor bed length for the parameter estimation is not recommended as the concentration gradient and the temperature profile would affect the intrinsic kinetic rates. However, it would be necessary for the validation and scale-up provided the incorporation of pressure drop and heat transfer formulations. In addition, employing more precise flow controllers and regulators would ensure consistent inlet gas composition and reduce fluctuations during different stages of the process. Collectively, these improvements would enhance the accuracy of transient concentration measurements and support more reliable kinetic parameter estimation.

## 5. Conclusion

This study developed a mechanistic kinetic model for a newly synthesised NiRu–Ca/Al DFM applied to cyclic $CO_2$ capture and methanation. A stage-wise characterisation of adsorption, purge, and hydrogenation was performed using a finite difference model that integrates transport, surface chemistry, and analyser delay effects.

An implicit finite difference scheme was used to simulate cyclic reactor behaviour, with a second-order response function incorporated to correct for analyser lag and tubing effects. This enabled accurate fitting of experimental concentration profiles, even for fast transients typically masked by system delays.

Bayesian optimisation (via Optuna) was employed for parameter estimation across different temperatures. A combined fitting strategy for purge and hydrogenation stages ensured mass continuity and helped infer purge kinetics based on downstream $CH_4$ trends. The model successfully reproduced observed behaviour and yielded physically consistent parameters.

Mechanistic analysis revealed that carbonate decomposition, not methanation, is rate-limiting during hydrogenation, explaining the low $CO_2$ breakthrough. A trade-off was also identified between adsorption capacity and methanation rate with temperature: while higher temperatures enhance reaction kinetics, they reduce $CO_2$ storage due to the exothermic nature of carbonation. As a result, maximum $CH_4$ yield occurred at 300 °C, highlighting the need for temperature optimisation.

Notably, the fitted hydrogenation kinetics deviated from those in steady-state Sabatier systems. The lower activation energy in the DFM process reflects the influence of spillover and carbonate decomposition unique to DFMs, supporting their promise for more efficient $CO_2$ conversion.

Beyond the kinetic findings, this work demonstrates a transparent, reproducible modelling pipeline using open-source tools, enabling future adaptation and peer validation. While the model captures key behaviours, early-stage adsorption and purge kinetics remain less resolved due to system delays. Reducing tubing lengths and improving flow control are recommended to enhance time resolution and fitting sensitivity.

These contributions form a robust foundation for advancing DFM-based processes, combining experimentally derived parameters for a newly formulated material with mechanistic insights into the interaction between temperature, kinetics, and reactor behaviour. The developed framework supports future efforts in optimisation, scale-up, and techno-economic evaluation of integrated ICCU.

## CRediT authorship contribution statement

Conceptualization, M.S. and M.S.D.; Methodology, M.S., M.S.D., M.D., A.D.W.; Software, M.S., M.D. and A.D.W.; Formal Analysis, A.D.W.; Investigation, M.D., S.B.G and L.P.M; Writing – Original Draft, M.D.; Writing – Review & Editing, M.S., M.S.D., A.D.W., S.B.G; Funding Acquisition, M.S. and M.S.D.; Resources, M.S., M.S.D., S.B,G, L.P.M; Supervision, M.S. and M.S.D.

## Declaration of competing interest

The authors declare that they have no known competing financial interests or personal relationships that could have appeared to influence the work reported in this paper.

## Data Availability

The code and datasets used in this study are available at Zenodo (DOI: 10.5281/zenodo.15678772) and maintained at GitHub (MeshkatD/DFM_Kinetics_PE).

## Acknowledgements

Financial support for this work was provided by the School of Chemistry and Chemical Engineering (Chemical and Process Engineering) and the Faculty of Engineering and Physical Sciences. This work was partially funded by the Engineering and Physical Sciences Research Council grant EP/X000753/1.

For the purpose of open access, the author has applied a Creative Commons attribution license (CC BY) to any Author Accepted Manuscript version arising from this submission